\documentclass[11pt, a4paper]{article}

\usepackage{cite}
\usepackage{amsmath}
\usepackage{amsfonts}
\usepackage{amssymb}
\usepackage{epsfig}

\begin{document}
\title{\bf{Lagrangian generators of the Poincar\'{e} gauge symmetries}}
\author{
{\bf {\normalsize Rabin Banerjee}}\thanks{Email: rabin@bose.res.in} \and {\bf {\normalsize Debraj Roy}}\thanks{E-mail: debraj@bose.res.in}\\
\and {\normalsize S.~N.~Bose National Centre for Basic Sciences,}\\
 {\normalsize Block-JD, Sector III, Salt Lake, Kolkata-700098, India.}\\
\and {\bf {\normalsize Saurav Samanta}}\thanks{E-mail: srvsmnt@gmail.com}\\
 {\normalsize Narasinha Dutt College,}\\
 {\normalsize 129, Belilious Road, Howrah-711101, India.}\\
}
\vspace{-0.5truecm}

\date{}

\maketitle

\begin{abstract}
We have systematically computed the generators of the symmetries arising in Poincar\'{e} gauge theory formulation of gravity, both in 2+1 and 3+1 dimensions. This was done using a completely Lagrangian approach. The results are expected to be valid in any dimensions, as seen through lifting the results of the 2+1 dimensional example into the 3+1 dimensional one.
\end{abstract} 

\section{Introduction}
\label{Sec:Intro}

Symmetry plays an important role in our understanding of nature and is the primary ingredient that goes into shaping of physical laws. Symmetry dictates the dynamics in nature \cite{Gross:1996}. This simple \emph{and} powerful insight has shaped our current knowledge of physics, right from the Galilean invariance of Newtonian mechanics, through the Lorentz invariance of special relativity, to the reparametrization or diffeomorphism invariance of general relativity. So understanding the role and systematic construction of symmetries through their generators is an important aspect in the study of any physical theory.

Given an action, a symmetry is some transformation of the fields of the theory involving arbitrary functions of time that leaves the action off-shell invariant. Rotations and translations in global space-time consist two important examples of such symmetries that are observed in nature. Together, these operations constitute the Poincar\'{e} group. To implement this Poincar\'{e} symmetry at the local level, whereby arbitrary functions of time and space are involved in the variations, new fields must be introduced which compensate for terms in the action not remaining invariant under this localization. These additional fields then are shown to represent the gravitational fields and the theory which describes this procedure and the corresponding dynamics is known as the Poincar\'{e} gauge theory (hereafter referred to as PGT) \cite{Utiyama:1956sy, Kibble:1961ba, Sciama:1962}. A lot of work has been done since then till recent times studying various aspects of the theory.\footnote{See \cite{Hehl:1976kj, Blagojevic:2002du} for reviews and further references.}

\begin{figure}[ht]

\label{Fig1}
\centerline{\epsfig{figure=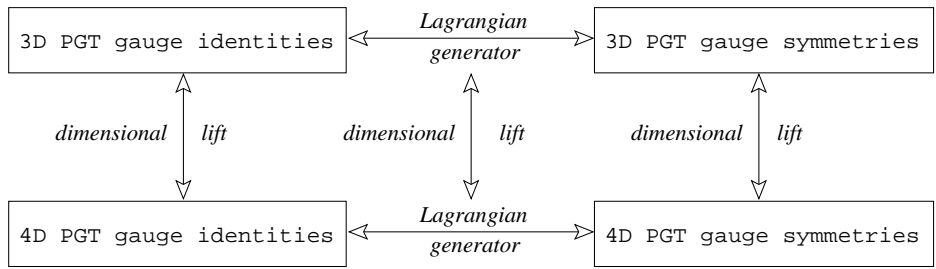,width=0.75\textwidth}}
\caption{\it{\ The scheme for Lagrangian analysis of Poincar\'{e} gauge theory (PGT):- symmetries, corresponding gauge identities and generators in 2+1 and 3+1 dimensions.}}

\end{figure}

The symmetries involved in PGT are motivated and derived by considering changes occurring due to local Lorentz rotations and infinitesimal coordinate changes (i.e. diffeomorphism). However, no canonical procedure of generating this symmetry through a set of generators exists till now. This problem was initially addressed in \cite{Blagojevic:2003uc,Blagojevic:2004hj} by computing the Hamiltonian (Gauss) generator following the approach of \cite{Castellani:1981us}. However the off-shell symmetries that were obtained as a result, were different from the PGT symmetries. The two sets matched only on-shell. This mismatch was thought to be a consequence of the approach \cite{Castellani:1981us} which, strictly speaking, is not a completely off-shell approach. Recently, an attempt to remedy this situation was given in \cite{Banerjee:2009vf}. A systematic and completely off-shell analysis of this issue was done in 2+1 dimensions, taking the 3D gravity model with torsion (modelled on the Mielke-Baekler action \cite{Mielke:1991nn}) and the Dirac Hamiltonian generator was computed. Alas, the original conclusions of \cite{Blagojevic:2003uc,Blagojevic:2004hj} remained unaltered. Here in this article, we adopt a totally different (Lagrangian) approach and show how to systematically construct generators of the PGT symmetries. The symmetries obtained from these generators reproduce the PGT symmetries without using any equations of motion.

The task of finding the symmetries of a given action is, in general, not trivial. They cannot always be found on inspection, as is possible, say, in electrodynamics. There exist two approaches to systematically construct the symmetries inbuilt in a given action; the Hamiltonian approach and the Lagrangian approach. In the Hamiltonian approach \cite{Castellani:1981us, Banerjee:2009vf, Henneaux:1990au, Banerjee:1999hu, Banerjee:1999yc, Banerjee:1999sz, Frolov:2009wu, Kiriushcheva:2006sg}, one first undertakes the Dirac classification of all constraints into first-class and second-class. Then the generators of the symmetries are obtained as some suitable combination of first-class constraints containing time derivatives of arbitrary functions. The Poisson bracket of the fields with these generators give us the symmetries as transformations. The Lagrangian method \cite{Mukunda:1974dr, Gitman:1990qh, Chaichian:1994ug, Shirzad:1998af, Samanta:2007fk, Banerjee:2009gr, Banerjee:2006jy}, on the other hand, hinges on the condition that the existence of symmetries necessarily implies the existence of certain identities involving quantities given by variations of the action w.r.t. the basic fields -- the Euler derivatives. The Lagrangian generators can then be found through comparison with the general expressions of such identities derived from a theory involving general symmetry transformation of fields in terms of arbitrary functions of time. We have used this later method of constructing generators here, after modifying it suitably for our model with vielbeins and connections as basic fields, that are written both in holonomic (global coordinate) and an-holonomic (local coordinate) indices.

Having constructed the generators first in 2+1 dimensions, we then repeat the process in 3+1 dimensions. The construction of symmetries through the Hamiltonian approach in 3+1 dimensions itself is a very difficult task and pure Dirac analysis and classification of constraints is non-trivial for, say, just the Einstein-Cartan theory. However it is shown that the Lagrangian version can be carried out more easily, leading to a systematic derivation of the symmetries even in 3+1 dimensions. To do this, we first lift the gauge identities corresponding to PGT symmetries from 2+1 to 3+1 dimensions and then carry out the Lagrangian analysis. However, at the end of the day, the symmetries obtained are the same as that found by directly lifting the 2+1 PGT symmetries in 3+1 dimensions. This shows that the PGT symmetries, arising out of geometrical considerations of reparametrization and local Lorentz symmetry, are beautifully consistent and useful in an extended sense.

The organization of the paper is as follows. In section \ref{Sec:GenForm} we give a general formulation of the Lagrangian method to find out generators, starting from known gauge identities of a model. Next, in section \ref{Sec:2+1LagGen}, we identify the gauge identities, calculate the generators in 2+1 dimensions and thus systematically reproduce the PGT symmetries in a Mielke-Baekler type model in 2+1 dimensions. For finding symmetry generators in 3+1 dimensions, we then lift the 2+1 dimensional gauge identities to 3+1 in section \ref{Sec:3+1LagGen} and find out the Lagrangian generators. These are shown to generate the usual PGT symmetries in 3+1 dimensions. Subsequently, we also demonstrate that the 2+1 dimensional generators, when stripped of their dimension dependant form containing dual fields, have the same form as in 3+1 dimensions. This suggests that our results may be extrapolated to any higher dimensions. The whole scheme is summarized in Figure \ref{Fig1}. Our conclusions are given in section \ref{Sec:Conclu}. Besides this, we have also included \ref{Sec:AppUsesLagGen}, where we comment on other possible applications of Lagrangian generators.

\paragraph{Conventions:}The coordinate frame or holonomic indices are written in Greek while Latin indices refer to the an-holonomic local Lorentz frame. The time and space bifurcation is indicated by choosing the beginning letters of both Greek $\left( \alpha, \beta, \ldots \right)$ and Latin $\left(a, b, \ldots\right)$ indices to run over the space indices, i.e. $1, 2, \ldots$ and choosing the middle letters of both Greek $\left( \mu, \nu, \ldots \right)$ and Latin $\left( i, j, \ldots \right)$ indices to run over both time and space indices $0, 1, 2, \ldots$. The totally antisymmetric tensor densities $\epsilon^{ijk}, \epsilon^{ijkl}, \epsilon^{\mu\nu\lambda} \text{ and } \epsilon^{\mu\nu\lambda\rho}$ are all normalized so that $\epsilon^{012} \text{ and } \epsilon^{0123}$ are unity. The spacetime signature chosen is $\left(+, -, -, \ldots \right)$. In specifying spacetime points, we have denoted by `$x$' both the time and space parts together while `$\bf{x}$' indicates only the spatial part of `$x$'. Thus $x\equiv(\mathrm{t},\bf{x})$.

\section{Lagrangian formulation: Gauge identities and symmetry generators}
\label{Sec:GenForm}

The action functional for a typical first-order theory invariant under the PGT symmetries in 2+1 dimensions, may be written by taking the triad $b^i_{\ \mu}$ and connection $\omega^i_{\ \mu}$ as independent variables, while constructing the configuration space \cite{Kibble:1961ba}. So in general, we have the action functional
\begin{align}
\label{Gen2plus1Action}
S = \int \mathrm{d}^3x ~\mathcal{L}\left[b^i_{\ \mu}(x),\omega^i_{\ \mu}(x)\right].
\end{align}
Arbitrary variations of the basic fields give rise to variation of the action in the following form
\begin{eqnarray}
\label{2plus1GenActionVar}
\delta S=-\int \mathrm{d}^3x ~\left\lbrace\,L_i^{\ \mu}(x) \,\delta b^i_\mu(x) + \bar{L}_i^{\ \mu}(x) \,\delta\omega^i_{\ \mu}(x) \,\right\rbrace
\end{eqnarray}
where $L_i^{\ \mu}:=-\dfrac{\delta\mathcal{L}}{\delta b^i_{\ \mu}}$ and $\bar{L}_i^{\ \mu}:=-\dfrac{\delta\mathcal{L}}{\delta\omega^i_{\ \mu}}$ are the Euler derivatives. The corresponding equations of motion are just
\begin{eqnarray}
\label{GenEqnsMotion}
L_i^{\ \mu}(x)=0 \qquad\text{and}\qquad \bar{L}_i^{\ \mu}(x)=0.
\end{eqnarray}
Now let us propose the following symmetries of the fields
\begin{align}
\label{GenVarFields}
\begin{aligned}
\delta b^i_{\ \mu}(x)=\displaystyle\sum_{s=0}^n (-1)^s\int \mathrm{d}^2{\bf z}~\left[\frac{\partial^s\xi^\sigma(z)}{\partial \mathrm{t}^s}\rho^i_{\ \mu\sigma(s)}(x,z) + \frac{\partial^s\theta^k(z)}{\partial \mathrm{t}^s}\zeta^i_{\ \mu k(s)}(x,z) \right]\\
\delta \omega^i_{\ \mu}(x)=\displaystyle\sum_{s=0}^n (-1)^s\int \mathrm{d}^2{\bf z}~\left[\frac{\partial^s\xi^\sigma(z)}{\partial \mathrm{t}^s}\bar{\rho}^i_{\ \mu\sigma(s)}(x,z) + \frac{\partial^s\theta^k(z)}{\partial \mathrm{t}^s}\bar{\zeta}^i_{\ \mu k(s)}(x,z) \right]
\end{aligned}
\end{align}
where the functions $\rho$, $\bar{\rho}$, $\zeta$ and $\bar{\zeta}$ are known as `Lagrangian generators' that generate variations in the basic fields while the six quantities $\xi^\sigma(z)$, $\theta^k(z)$ are functions of space and time serving as infinitesimal gauge parameters. Note that their number is governed by the Poincar\'{e} symmetry group, which in 2+1 dimensions has six independent symmetries. These variations of fields are symmetries in the sense that $S\left[b^i_{\ \mu}+\delta b^i_{\ \mu},\ \omega^i_\mu + \delta\omega^i_{\ \mu}\right]=S\left[b^i_{\ \mu},\omega^i_{\ \mu}\right]$, or equivalently $\delta S=0$ under these variations.

Substituting the variations \eqref{GenVarFields} of $b$ and $\omega$ in \eqref{2plus1GenActionVar} yields the variation of the action as
\begin{align}
\label{ex b1}
\begin{aligned}
\delta S &=& \!\!-\int \!\!\mathrm{d}^2{\bf x} \int \!\!\mathrm{d}^2{\bf z} \,\displaystyle\sum_{s=0}^n (-1)^s \!\int \!\!\mathrm{d}\mathrm{t} \left[\frac{\partial^s\xi^\sigma(z)}{\partial \mathrm{t}^s} \rho^i_{\ \mu\sigma(s)}(x,z) L_i^{\ \mu}(x) + \frac{\partial^s\theta^k(z)}{\partial \mathrm{t}^s} \zeta^i_{\ \mu k(s)}(x,z) L_i^{\ \mu}(x)\right]\\
&\ & \!\!-\int \!\!\mathrm{d}^2{\bf x} \int \!\!\mathrm{d}^2{\bf z} \,\displaystyle\sum_{s=0}^n (-1)^s \!\int \!\!\mathrm{d}\mathrm{t} \left[\frac{\partial^s\xi^\sigma(z)}{\partial \mathrm{t}^s} \bar{\rho}^i_{\ \mu\sigma(s)}(x,z) \bar{L}_i^{\ \mu}(x) + \frac{\partial^s\theta^k(z)}{\partial \mathrm{t}^s} \bar{\zeta}^i_{\ \mu k(s)}(x,z) \bar{L}_i^{\ \mu}(x)\right].
\end{aligned}
\end{align}
To simplify the above expression, let us take a representative term -- the first term in the first line -- from \eqref{ex b1}. The other terms, having similar structure, can then be handled by a similar technique.
\begin{align}
\begin{aligned}
-& \int \!\!\mathrm{d}^2{\bf x} \int \!\!\mathrm{d}^2{\bf z} \,\displaystyle\sum_{s=0}^n (-1)^s \int \!\!\mathrm{d}\mathrm{t} ~\frac{\partial^s\xi^\sigma(z)}{\partial \mathrm{t}^s} \rho^i_{\ \mu\sigma(s)}(x,z) L_i^{\ \mu}(x)\\
= & \!\! \;-\int \!\!\mathrm{d}^2{\bf x} \int \!\!\mathrm{d}^2{\bf z} \,\left[\int \!\!\mathrm{d}\mathrm{t} ~\xi^\sigma(z) \rho^i_{\ \mu\sigma(0)}(x,z) L_i^{\ \mu}(x) - \int \!\!\mathrm{d}\mathrm{t}~\frac{\partial\xi^\sigma(z)}{\partial \mathrm{t}} \rho^i_{\ \mu\sigma(1)}(x,z) L_i^{\ \mu}(x) + \ldots \right.\\
& \left. \qquad\qquad\qquad\qquad \ldots + (-1)^n\int \!\!\mathrm{d}\mathrm{t} ~\frac{\partial^n\xi^\sigma(z)}{\partial \mathrm{t}^n} \rho^i_{\ \mu\sigma(n)}(x,z) L_i^{\ \mu}(x) \right].
\end{aligned}
\end{align}
We now interchange derivatives (wherever applicable) by using partial integrals, throwing away the boundary terms by assuming the fields to be well behaved at infinity. Also note that we have precisely the same number of negative signs before each term through the $(-1)^s$ factor, as required for carrying out partial integrals `$s$' times. So, simplifying the above equation yields
\begin{align}
\begin{aligned}
& -\int \!\!\mathrm{d}^2{\bf x} \int \!\!\mathrm{d}^2{\bf z} \,\left[\int \!\!\mathrm{d}\mathrm{t} ~\xi^\sigma(z) \left\lbrace\rho^i_{\ \mu\sigma(0)}(x,z) L_i^{\ \mu}(x)\right\rbrace + \int \!\!\mathrm{d}\mathrm{t}~\xi^\sigma(z)\frac{\partial}{\partial \mathrm{t}} \left\lbrace\rho^i_{\ \mu\sigma(1)}(x,z) L_i^{\ \mu}(x) \right\rbrace+ \right.\\
& \left. \qquad\qquad\qquad\qquad \ldots + \int \!\!\mathrm{d}\mathrm{t} ~\xi^\sigma(z)~\frac{\partial^n}{\partial \mathrm{t}^n} \left\lbrace\rho^i_{\ \mu\sigma(n)}(x,z) L_i^{\ \mu}(x) \right\rbrace\right]\\
& \!\!= \;-\int \!\!\mathrm{d}^2{\bf z} \int \!\!\mathrm{d}\mathrm{t} ~\xi^\sigma(z)~\left[\displaystyle \sum_{s=0}^n\int \mathrm{d}^2{\bf x} ~\frac{\partial^s}{\partial\mathrm{t}^s} \left\lbrace \rho^i_{\ \mu\sigma(s)}(x,z) L_i^{\ \mu}(x) \right\rbrace\right].
\end{aligned}
\end{align}
Substituting this back in the variation of the action \eqref{ex b1}, we get
\begin{align}
\delta S = -\int \mathrm{d}^3z \left[ \xi^\sigma(z) \,\Lambda_\sigma(z) + \theta^k(z) \,\Xi_k(z)\right]
\end{align}
where $\Lambda$ and $\Xi$ are defined as:
\begin{align}
\begin{aligned}
\label{systemGaugeId}
\Lambda_\sigma(z) &=& \displaystyle\sum_{s=0}^n \int \mathrm{d}^2{\bf x} ~\frac{\partial^s}{\partial \mathrm{t}^s} \left\lbrace \rho^i_{\ \mu\sigma(s)}(x,z)L_i^{\ \mu}(x) + \bar{\rho}^i_{\ \mu\sigma(s)}(x,z)\bar{L}_i^{\ \mu}(x) \right\rbrace\\
\Xi_k(z) &=& \displaystyle\sum_{s=0}^n \int \mathrm{d}^2{\bf x} ~\frac{\partial^s}{\partial \mathrm{t}^s} \left\lbrace \zeta^i_{\ \mu k(s)}(x,z)L_i^{\ \mu}(x) + \bar{\zeta}^i_{\ \mu k(s)}(x,z)\bar{L}_i^{\ \mu}(x) \right\rbrace.\\
\end{aligned}
\end{align}
Since each of the gauge parameters $\xi^\sigma$ and $\theta^k$ are independent quantities, the invariance of the action $\left(\delta S=0\right)$ implies the following conditions
\begin{align}
\label{shortGaugeIds}
\begin{aligned}
\Lambda_\sigma(z) &= 0\\
\Xi_k(z) &= 0.
\end{aligned}
\end{align}
These are known as `\emph{gauge identities}'. They are identities in the sense that on substituting the Euler derivatives $L_i^{\ \mu}$ and $\bar{L}_i^{\ \mu}$ in \eqref{systemGaugeId}, all terms cancel out and we see that the relations are zero identically. Note that until now we have only used the definition of the Euler derivatives in terms of variation of fields $\left( \text{for example } L_i^{\ \mu}=-\dfrac{\delta \mathcal{L}}{\delta b^i_{\ \mu}}~\right)$, but we have not set the Euler derivative to zero, i.e. we have not used any equations of motion. In fact, using the equations of motion trivializes the gauge identities as $0=0$ relations. 

Now the algorithm for finding out the Lagrangian symmetry generators is simple. Given an action, we can easily find the Euler derivatives by varying the action w.r.t. the basic fields. Using the Euler derivatives, we can then try to build a set of independent, identically vanishing equations - the gauge identities. Alternatively, we can also explicitly check any set of identities proposed to hold as gauge identities from physical considerations. The number of these gauge identities is identical to the number of independent symmetries. In this particular case of PGT, they are six in number and are stated compactly in \eqref{shortGaugeIds}. Once we obtain a set of \emph{independent} gauge identities for the action, we then finally compare the given identities with the general form presented in \eqref{systemGaugeId}, and find out the generators $\left(\text{denoted here as } \rho, ~\bar{\rho}, ~\zeta \text{ and } ~\bar{\zeta} \,\right)$.

\section{Lagrangian generators for 2+1 dimensional PGT symmetries}
\label{Sec:2+1LagGen}

Gravity in 2+1 dimensions is widely studied, both as a simplified problem in comparison to 3+1 dimensions, and also as a field which is interesting in its own right. Among various models studied, one is the Mielke-Baekler model which explicitly includes a torsion term, along with the Chern-Simons action and the usual Einstein-Cartan piece. The particular Mielke-Baekler type action of recent interest \cite{Banerjee:2009vf, Blagojevic:2009xw, Blagojevic:2003uc, Blagojevic:2004hj, Basu:2009dy} that we study here is:
\begin{align}
\label{EC2plus1}
S=\int \mathrm{d}^3x ~\epsilon^{\mu\nu\rho}\left[ab^i_{\ \mu} R_{i\nu\rho}-\frac{\Lambda}{3} \epsilon_{ijk}b^i_{\ \mu} b^j_{\ \nu} b^k_{\ \rho} + \alpha_3 \left(\omega^i_{\ \mu} \partial_\nu\omega_{i\rho} + \frac{1}{3} \epsilon_{ijk}\omega^i_{\ \mu} \omega^j_{\ \nu} \omega^k_{\ \rho} \right) + \frac{\alpha_4}{2} b^i_{\ \mu} T_{i\nu\rho} \right],
\end{align}
where $a$, $\Lambda$, $\alpha_3$ and $\alpha_4$ are arbitrary parameters, $R_{i\nu\rho}$ is the curvature tensor defined as
\begin{align}
\label{2plus1Riemann}
R_{i\nu\rho}=\partial_\nu \omega_{i\rho} - \partial_\rho \omega_{i\nu} + \epsilon_{ijk}\omega^j_\nu\omega^k_\rho,
\end{align}
and $T_{i\nu\rho}$ is the torsion given by,
\begin{align}
\label{2+1Torsion}
T_{i\nu\rho}=\partial_\nu b_{i\rho} - \partial_\rho b_{i\nu} + \epsilon_{ijk}\omega^j_{\ \nu}b^k_{\ \rho} - \epsilon_{ijk}\omega^j_{\ \rho}b^k_{\ \nu}.
\end{align}
The first term proportional to `$a$' is the Einstein-Hilbert action written in three dimensions using the identity
\begin{align*}
bR = -\epsilon^{\mu\nu\rho}\,b^i_{\ \mu} R_{i\nu\rho}
\end{align*}
where $b = \text{det}\,b^i_{\ \mu}$ and $R = b_i^{\ \mu} b_j^{\ \nu} R^{ij}_{\ \ \mu\nu}$. The second term is the cosmological constant part, the third one is the Chern-Simons action while the fourth includes torsion. These terms can be manipulated with the help of the adjustable parameters $a$, $\Lambda$, $\alpha_3$ and $\alpha_4$.

The action \eqref{EC2plus1} is known to be invariant under the following PGT symmetries \cite{Banerjee:2009vf,Blagojevic:2003uc,Blagojevic:2004hj}
\begin{align}
\begin{aligned}
\label{fieldtrans3D}
\delta b^i_{\ \mu} &= -\epsilon^i_{\ jk}b^j_{\ \mu}\theta^k - \partial_\mu\xi^\rho \, b^i_{\ \rho} - \xi^{\rho}\,\partial_{\rho}b^i_{\ \mu}\\
\delta \omega^{i}_{\ \mu} &= -\left(\partial_\mu\theta^i + \epsilon^i_{\ jk}\omega^j_{\ \mu}\theta^k \right) - \partial_\mu\xi^\rho \, \omega^i_{\ \rho} - \xi^{\rho}\,\partial_{\rho}\omega^i_{\ \mu}.
\end{aligned}
\end{align}

A set of independent gauge identities corresponding to PGT symmetries for the action \eqref{EC2plus1} is already known \cite{Banerjee:2009vf}
\begin{align}
\begin{aligned}
\label{prevGaugeId}
\Lambda_\sigma &:= -L_i^{\ \mu} \partial_\sigma b^i_{\ \mu} - \bar{L}_i^{\ \mu} \partial_\sigma \omega^i_{\ \mu} + \partial_\mu\left(b^i_{\ \sigma}L_i^{\ \mu}\right) + \partial_\mu\left(\omega^i_{\ \sigma}\bar{L}_i^{\ \mu}\right)=0\\
\Xi_k &:= -\epsilon^i_{\ jk}L_i^{\ \mu}b^j_{\ \mu} - \epsilon^i_{\ jk}\bar{L}_i^{\ \mu}\omega^j_{\ \mu} + \partial_\mu \bar{L}_k^{\ \mu}=0.
\end{aligned}
\end{align}
Here $L_i^{\ \mu}$ and $\bar{L}_i^{\ \mu}$ are the Euler derivatives obtained from the action \eqref{EC2plus1}, and are given by,
\begin{align}
\label{Euler2plus1}
\begin{aligned}
L_i^{\ \mu} &:= -\frac{\delta \mathcal{L}}{\delta b^i_{\ \mu}} = -\epsilon^{\mu\nu\rho}\left[a R_{i\nu\rho} +\alpha_4 T_{i\nu\rho} - \Lambda\epsilon_{ijk}b^j_{\ \nu}b^k_{\ \rho}\right]\\
\bar{L}_i^{\ \mu} &:= -\frac{\delta \mathcal{L}}{\delta \omega^i_{\ \mu}} = -\epsilon^{\mu\nu\rho}\left[\alpha_3R_{i\nu\rho} + aT_{i\nu\rho} - \Lambda\epsilon_{ijk}b^j_{\ \nu}b^k_{\ \rho}\right].
\end{aligned}
\end{align}
Substituting these in \eqref{prevGaugeId} it may easily be checked that all terms cancel and they are indeed identities. Now the Lagrangian symmetry generators which will give us a set of symmetries of the action \eqref{EC2plus1} may be found by comparing these identities \eqref{prevGaugeId} with the general gauge identities derived before in \eqref{systemGaugeId}.

We have employed the following strategy in comparing the two relations in question. Any sum over Greek (holonomic) indices is broken into the time and space part, {\it i.e.} say, $A^\mu B_\mu = A^0 B_0 + A^\alpha B_\alpha$; the gauge identity $\Lambda_\sigma$ is also broken into sets $\Lambda_0$ and $\Lambda_\alpha$; and finally coefficients (in general field dependant) of the Euler derivatives $L_i^{\ 0}$, $L_i^{\ \beta}$, etc. are matched between the two relations \eqref{systemGaugeId} and \eqref{prevGaugeId}.

Let us now illustrate the details for one particular term: coefficient of $L_i^{\ \beta}$ in $\Lambda_\alpha$. An inspection of \eqref{prevGaugeId} reveals that there occur terms either with zero or a single time derivative. This implies that the summation over `$s$' in \eqref{systemGaugeId} is restricted to only two values, $s=0,\, 1.$ So, the relevant terms from \eqref{systemGaugeId} are,
\begin{align}
\label{ex a1}
\int ~\mathrm{d}^2{\bf x} ~\left\lbrace \rho^i_{\ \beta\alpha(0)}(x,z)\,L_i^{\ \beta}(x) + \partial_0\rho^i_{\ \beta\alpha(1)}(x,z)\,L_i^{\ \beta}(x) + \rho^i_{\ \beta\alpha(1)}(x,z) \,\partial_0 L_i^{\ \beta}(x) \right\rbrace
\end{align}
while those from \eqref{prevGaugeId} are
\begin{align*}
-&L_i^{\ \beta}(z)\,\partial_\alpha b^i_{\ \beta}(z) + \partial_\beta \left(b^i_{\ \alpha}(z)\,L_i^{\ \beta}(z)\right).
\end{align*}
The above expression may be recast in the form
\begin{align}
\label{ex a2}
\int \mathrm{d}^2{\bf x} ~\Big\lbrace -&L_i^{\ \beta}(x)\,\partial_\alpha b^i_{\ \beta}(x)\,\delta({\bf x}-{\bf z}) - b^i_{\ \alpha}(x)\,L_i^{\ \beta}(x)~\partial_\beta^{(\bf x)}\delta({\bf x}-{\bf z}) ~\Big\rbrace\,,
\end{align}
to facilitate comparison with \eqref{ex a1}. This comparison yields the generators $\rho^i_{\ \beta\alpha(1)}=0$ and $\rho^i_{\ \beta\alpha(0)} = -\partial_\alpha b^i_{\ \beta}(x) ~\delta({\bf x}-{\bf z}) - b^i_{\ \alpha}(x)~\partial_\beta^{(\bf x)}\delta({\bf x}-{\bf z})$. The other generators may also be found in a similar manner. We list all the non-zero ones below:
\begin{align}
\label{2plus1GenSet1}
\begin{aligned}
\rho^i_{\ 0\sigma(0)}(x,z) &= -\partial_\sigma b^i_{\ 0}(x) ~\delta({\bf x}-{\bf z})\\
\rho^i_{\ \alpha \sigma(0)}(x,z) &= -\partial_\sigma b^i_{\ \alpha}(x) ~\delta({\bf x}-{\bf z}) - b^i_{\ \sigma}(x)~\partial_\alpha^{(\bf x)}\delta({\bf x}-{\bf z})\\
\rho^i_{\ 0\sigma(1)}(x,z) &= {\ } b^i_{\ \sigma}(x) ~\delta({\bf x}-{\bf z}),\\
\end{aligned}
\end{align}
\begin{align}
\label{2plus1GenSet2}
\begin{aligned}
\bar{\rho}^i_{\ 0\sigma(0)}(x,z) &= -\partial_\sigma \omega^i_{\ 0}(x) ~\delta({\bf x}-{\bf z})\\
\bar{\rho}^i_{\ \alpha\sigma(0)}(x,z) &= -\partial_\sigma \omega^i_{\ \alpha}(x) ~\delta({\bf x}-{\bf z}) - \omega^i_{\ \sigma}(x)~\partial_\alpha^{(\bf x)}\delta({\bf x}-{\bf z})\\
\bar{\rho}^i_{\ 0\sigma(1)}(x,z) &= {\ } \omega^i_{\ \sigma}(x) ~\delta({\bf x}-{\bf z}),\\
\end{aligned}
\end{align}
\begin{align}
\label{2plus1GenSet3}
\begin{aligned}
\zeta^i_{\ \sigma k(0)}(x,z) &= -\epsilon^i_{\ jk} b^j_{\ \sigma}(x) ~\delta({\bf x}-{\bf z}),\\
\end{aligned}
\end{align}
\begin{align}
\label{2plus1GenSet4}
\begin{aligned}
\bar{\zeta}^i_{\ 0k(0)}(x,z) &= -\epsilon^i_{\ jk} \omega^j_{\ 0}(x) ~\delta({\bf x}-{\bf z})\\
\bar{\zeta}^i_{\ \alpha k(0)}(x,z) &= -\epsilon^i_{\ jk} \omega^j_{\ \alpha}(x) ~\delta({\bf x}-{\bf z}) - \delta^i_k~\partial_\alpha^{({\bf x})} \delta({\bf x}-{\bf z})\\
\bar{\zeta}^i_{\ 0k(1)}(x,z) &= {\ } \delta^i_k ~\delta({\bf x}-{\bf z}).\\
\end{aligned}
\end{align}
Thus, having obtained all the Lagrangian generators, it is now possible to generate the transformations of the basic fields $b$ and $\omega$ through \eqref{GenVarFields}. We illustrate the process for $b^i_{\ \alpha}$.
\begin{align}
\begin{aligned}
\delta b^i_{\ \alpha}(x)=&\int \mathrm{d}^2{\bf z} ~\left[ \xi^0(z)\,\rho^i_{\ \alpha 0(0)}(x,z) + \xi^\beta(z)\,\rho^i_{\ \alpha\beta(0)}(x,z) + \theta^k(z)\,\zeta^i_{\ \alpha k(0)}(x,z) \right]\\
& \qquad-\int \mathrm{d}^2{\bf z} ~\left[ \partial_0\xi^0(z)\,\rho^i_{\ \alpha 0(1)}(x,z) + \partial_0\xi^\beta(z)\,\rho^i_{\ \alpha\beta(1)}(x,z) + \partial_0\theta^k(z)\,\zeta^i_{\ \alpha k(1)}(x,z) \right]
\end{aligned}
\end{align}
Using the form of the generators $\rho,\,\zeta$ given in \eqref{2plus1GenSet1} and \eqref{2plus1GenSet3}, one obtains,
\begin{align}
\begin{aligned}
\delta b^i_{\ \alpha}(x)=&\int \mathrm{d}^2{\bf z} ~\left[ \xi^0(z)\,\left\lbrace -\partial_0 b^i_{\ \alpha}(x)\,\delta({\bf x}-{\bf z}) - b^i_{\ 0}(x)~\partial_\alpha^{(\bf x)}\delta({\bf x}-{\bf z}) \right\rbrace \right.\\
&\left. \qquad\;\: + \;\xi^\beta(z)\,\left\lbrace -\partial_\beta b^i_{\ \alpha}(x) ~\delta({\bf x}-{\bf z}) - b^i_{\ \beta}(x)~\partial_\alpha^{(\bf x)}\delta({\bf x}-{\bf z}) \right\rbrace  \right.\\
&\left. \qquad\;\: + \;\theta^k(z)\,\left\lbrace -\epsilon^i_{\ jk} b^j_{\ \alpha}(x) ~\delta({\bf x}-{\bf z}) \right\rbrace \right] \ - \ \int \mathrm{d}^2{\bf z} ~\left[\ 0 + 0 + 0\ \right]\\
=& -\epsilon^i_{\ jk}\,b^j_{\ \alpha}\,\theta^k - \partial_\alpha\xi^\mu\,b^i_{\ \mu} - \xi^\mu\,\partial_\mu b^i_\alpha,
\end{aligned}
\end{align}
which corresponds to the $\mu=\alpha$ (space) component of the PGT symmetry $\delta b^i_{\ \mu}$ given in \eqref{fieldtrans3D}. All other PGT symmetries given in \eqref{fieldtrans3D} are easily reproduced by this procedure.

\section{PGT construction in 3+1 dimensions and Lagrangian analysis}
\label{Sec:3+1LagGen}

The same PGT symmetries, being constructed out of local Lorentz and general diffeomorphism symmetries, are respected by a wide class of Lagrangians \cite{Blagojevic:2002du}. In 3+1 dimensions, the Chern-Simons term of 2+1 dimensions \eqref{EC2plus1} automatically drops out. The other terms have their counterparts in 3+1 dimensions, in addition to some other new possible terms. However, for simplicity, it suffices our aim of constructing the generators of PGT symmetries, to consider only the most important part of the gravitational action -- the Einstein-Cartan term. Thus we take the following action in 3+1 dimensions,
\begin{align}
\label{EC3plus1}
S=\int \mathrm{d}^4x ~b\,R
\end{align}
where $b=\text{det}\left(b^i_{\ \mu}\right)$ and the curvature scalar $R=b_i^{\ \mu}b_j^{\ \nu}R^{ij}_{\ \ \mu\nu}$. The curvature tensor and the torsion tensor are defined as:
\begin{align}
\label{3plus1CurvTor}
\begin{aligned}
T^i_{\ \mu\nu} &= \partial_\mu b^i_{\ \nu} - \partial_\nu b^i_{\ \mu} + \omega^{i}_{\ \, k\mu} b^k_{\ \nu} - \omega^{i}_{\ \,k\nu} b^k_{\ \mu}\\
R^{ij}_{\ \ \mu\nu} &= \partial_\mu \omega^{ij}_{\ \ \nu} - \partial_\nu \omega^{ij}_{\ \ \mu} + \omega^i_{\ k\mu}\omega^{kj}_{\ \ \nu} - \omega^i_{\ k\nu}\omega^{kj}_{\ \ \mu}.
\end{aligned}
\end{align}
The corresponding Euler derivatives can be found in the standard way
\begin{align}
\label{3plus1EulerDeriv}
\begin{aligned}
L_i^{\ \mu} &:= -\frac{\delta \mathcal{L}}{\delta b^i_{\ \mu}}  = -2b \left( R_i^{\ \mu} + \frac{1}{2} b_i^{\ \mu}R \right)\\
L_{ij}^{\ \ \mu} &:= -\frac{\delta \mathcal{L}}{\delta \omega^{ij}_{\ \ \mu}}  = b \left( \,b_s^{\ \mu}\,T^s_{\ ij} + b_i^{\ \mu}\,T^s_{\ js} - b_j^{\ \mu}\,T^s_{\ is} \,\right).
\end{aligned}
\end{align}

To find the appropriate gauge identities here, we will now take help of the identities found previously for 2+1 dimensions. The 2+1 dimensional model was constructed using the basic fields $b^i_{\ \mu}$ and $\omega^i_{\ \mu}$, where the latter was a dual construct of the field $\omega^{ij}_{\ \ \mu}$, valid only in 2+1 dimensions. We would now like to write the gauge identity \eqref{prevGaugeId} in terms of the fields $b^i_{\ \mu}$ and $\omega^{ij}_{\ \ \mu}$, thus getting rid of the use of special 2+1 dimensional properties. The resultant identities will then be proposed for 3+1 dimensions and a Lagrangian analysis will be carried out to find out the corresponding symmetries.

Now let us consider the $\Xi_k$ identity in \eqref{prevGaugeId}. Contracting it with the Levi-Civita symbol, we find,
\begin{align}
\label{EpsilonContr}
\Xi_{mn} = -\epsilon_{mn}^{\ \ \ k}\:\Xi_k = L_m^{\ \;\mu}\,b_{n\mu} - L_n^{\ \mu}\,b_{m\mu} + \bar{L}_m^{\ \mu}\,\omega_{n\mu} - \bar{L}_n^{\ \mu}\,\omega_{m\mu} - \epsilon_{mn}^{\ \ \ k}\,\partial_\mu \bar{L}_k^{\ \mu}.
\end{align}
Next, introducing relations between the dual fields and their corresponding counterparts through
\begin{align}
\label{LiftMaps}
\begin{aligned}
\omega^i_{\ \mu} &= -\frac{1}{2} \,\epsilon^i_{\ jk}\,\omega^{jk}_{\ \ \mu}\\
\bar{L}_i^{\ \mu} &= -\epsilon_i^{\ jk} \, L_{jk}^{\ \ \mu},
\end{aligned}
\end{align}
and using the following identity for Levi-Civita symbols
\begin{align}
\label{LeviCivitaProduct}
\epsilon_m^{\ \ hp}\,\epsilon_{njk} = \eta_{mn}\delta^h_j\delta^p_k - \eta_{mn}\delta^h_k\delta^p_j - \eta_{mj}\delta^h_n\delta^p_k + \eta_{mj}\delta^h_k\delta^p_n + \eta_{mk}\delta^h_n\delta^p_j - \eta_{mk}\delta^h_j\delta^p_n,
\end{align}
we are able to write the gauge identity completely in terms of the fields $b^i_{\ \mu}$ and $\omega^{ij}_{\ \ \mu}$. The other gauge identity $\Lambda_\sigma$ in the set of the 2+1 dim identities \eqref{prevGaugeId} can also be ridden of the duals through a similar procedure. The resultant set of identities are:
\begin{align}
\label{3plus1PGTGaugeIds}
\begin{aligned}
\Lambda_\sigma &:= -L_i^{\ \mu}\partial_\sigma b^i_{\ \mu} - L_{ij}^{\ \ \mu}\partial_\sigma\omega^{ij}_{\ \ \mu} + \partial_\mu\left(L_i^{\ \mu}b^i_{\ \sigma}\right) + \partial_\mu\left(L_{ij}^{\ \ \mu}\omega^{ij}_{\ \ \sigma}\right)=0\\
\Xi_{ij} &:= {\ } L_{i\mu}\,b_j^{\ \mu} - L_{j\mu}\,b_i^{\ \mu} + 2\left(\partial_{\nu}L_{ij}^{\ \ \nu}-L_{ik}^{\ \ \nu}\omega^k_{\ j\nu} + L_{jk}^{\ \ \nu} \omega^k_{\ i\nu}\right)=0.
\end{aligned}
\end{align}
Since these are now written independent of any dimensionally dependant dual fields, we may propose that they also hold in 3+1 dimensions. An explicit check, using the expressions for the Euler derivatives \eqref{3plus1EulerDeriv} confirms the proposition.

Expressing the action \eqref{Gen2plus1Action} in terms of basic fields, rather than the duals, we have
\begin{align}
\label{Gen3plus1Action}
S=\int \,\mathrm{d}^4x~\mathcal{L}\left[ b^i_{\ \mu},\,\omega^{ij}_{\ \ \mu} \right].
\end{align}
The symmetry transformations of this action are now given by
\begin{align}
\label{3plus1GenVarFieldsPGT}
\begin{aligned}
\delta b^i_{\ \mu}(x) &= \displaystyle\sum_{s=0}^n (-1)^s\int \mathrm{d}^3{\bf z}~\left[\frac{\partial^s\xi^\sigma(z)}{\partial \mathrm{t}^s}\rho^i_{\ \mu\sigma(s)}(x,z) + \frac{\partial^s\theta^{lk}(z)}{\partial \mathrm{t}^s}\zeta^i_{\ \mu lk(s)}(x,z) \right]\\
\delta \omega^{ij}_{\ \ \mu}(x) &= \displaystyle\sum_{s=0}^n (-1)^s\int \mathrm{d}^3{\bf z}~\left[\frac{\partial^s\xi^\sigma(z)}{\partial \mathrm{t}^s}\bar{\rho}^{ij}_{\ \ \mu\sigma(s)}(x,z) + \frac{\partial^s\theta^{lk}(z)}{\partial \mathrm{t}^s}\bar{\zeta}^{ij}_{\ \ \mu lk(s)}(x,z) \right],
\end{aligned}
\end{align}
which are the 3+1 dimensional versions of \eqref{GenVarFields}. Now adopting identical steps as in section \ref{Sec:GenForm}, we obtain the analogues of the gauge identities \eqref{systemGaugeId} with $\Lambda_\sigma$ and $\Xi_{lk}$ given by,
\begin{align}
\label{3plus1PGTSytemGaugeIds}
\begin{aligned}
\Lambda_\sigma(z) &= \displaystyle\sum_{s=0}^n\int \textrm{d}^3{\bf{x}} ~\frac{\partial^s}{\partial \mathrm{t}^s} \left\lbrace \rho^i_{\ \mu\sigma(s)}(x,z)L_i^{\ \mu}(x) + \bar{\rho}^{ij}_{\ \ \mu\sigma(s)}(x,z) L_{ij}^{\ \ \mu}(x) \right\rbrace\\
\Xi_{lk}(z) &= \displaystyle\sum_{s=0}^n\int \textrm{d}^3{\bf{x}} ~\frac{\partial^s}{\partial \mathrm{t}^s} \left\lbrace \zeta^i_{\ \mu lk(s)}(x,z) L_i^{\ \mu}(x) + \bar{\zeta}^{ij}_{\ \ \mu lk(s)}(x,z) L_{ij}^{\ \ \mu}(x) \right\rbrace.
\end{aligned}
\end{align}

Next, we compare these with the set of identities \eqref{3plus1PGTGaugeIds} term by term as explained in the discussion above eq. \eqref{ex a1} to find out the relevant Lagrangian generators. The non-zero ones are listed below
\begin{align}
\label{3plus1GenSet1}
\begin{aligned}
\rho^i_{\ 0\sigma(0)}(x,z) &= -\partial_\sigma b^i_{\ 0}(x) ~\delta({\bf x}-{\bf z})\\
\rho^i_{\ \alpha\sigma(0)}(x,z) &= -\partial_\sigma b^i_{\ \alpha}(x) ~\delta({\bf x}-{\bf z}) - b^i_{\ \sigma}(x)~\partial_\alpha^{(\bf x)}\delta({\bf x}-{\bf z})\\
\rho^i_{\ 0\sigma(1)}(x,z) &= {\ } b^i_{\ \sigma}(x) ~\delta({\bf x}-{\bf z}),\\
\end{aligned}
\end{align}
\begin{align}
\label{3plus1GenSet2}
\begin{aligned}
\bar{\rho}^{ij}_{\ \ 0\sigma(0)}(x,z) &= -\partial_\sigma \omega^{ij}_{\ \ 0}(x) ~\delta({\bf x}-{\bf z})\\
\bar{\rho}^{ij}_{\ \ \alpha\sigma(0)}(x,z) &= -\partial_\sigma \omega^{ij}_{\ \ \alpha}(x) ~\delta({\bf x}-{\bf z}) - \omega^{ij}_{\ \ \sigma}(x)~\partial_\alpha^{(\bf x)}\delta({\bf x}-{\bf z})\\
\bar{\rho}^{ij}_{\ \ 0\sigma(1)}(x,z) &= {\ } \omega^{ij}_{\ \ \sigma}(x) ~\delta({\bf x}-{\bf z}),\\
\end{aligned}
\end{align}
\begin{align}
\label{3plus1GenSet3}
\begin{aligned}
\zeta^i_{\ \mu lk(0)}(x,z) &= {\ } \frac{1}{2}\left[\delta^i_l \,b_{k\mu}(x)-\delta^i_k \,b_{l\mu}(x)\right] ~\delta({\bf x}-{\bf z}),\\
\end{aligned}
\end{align}
\begin{align}
\label{3plus1GenSet4}
\begin{aligned}
\bar{\zeta}^{ij}_{\ \ 0lk(0)}(x,z) &= -\frac{1}{2}\left[ \delta^i_l\omega^j_{\ k0} - \delta^i_k\omega^j_{\ l0} - \delta^j_l\omega^i_{\ k0} + \delta^j_k\omega^i_{\ l0}  \right]\delta({\bf x}-{\bf z})\\
\bar{\zeta}^{ij}_{\ \ \alpha lk(0)}(x,z) &= -\frac{1}{2}\left[ \delta^i_l\omega^j_{\ k\alpha} - \delta^i_k\omega^j_{\ l\alpha} - \delta^j_l\omega^i_{\ k\alpha} + \delta^j_k\omega^i_{\ l\alpha}  \right]\delta({\bf x}-{\bf z}) - \frac{1}{2}\left[ \delta^i_l\delta^j_k - \delta^i_k\delta^j_l \right] \,\partial_\alpha^{(\bf x)}\delta({\bf x}-{\bf z})\\
\bar{\zeta}^{ij}_{\ \ 0lk(1)}(x,z) &= {\ \ } \frac{1}{2} \left[ \delta^i_l\delta^j_k - \delta^i_k\delta^j_l \right] ~\delta({\bf x}-{\bf z}).\\
\end{aligned}
\end{align}
These generators will yield the symmetries of the action \eqref{EC3plus1} through the transformation \eqref{3plus1GenVarFieldsPGT}. An explicit calculation, leads to the symmetries
\begin{align}
\label{3plus1PGTFieldTrans}
\begin{aligned}
\delta b^i_{\ \mu} &= \theta^i_{\ k} b^k_{\ \mu} - \partial_\mu\xi^\rho b^i_{\ \rho} - \xi^{\rho}\partial_{\rho} b^i_{\ \mu}\\
\delta \omega^{ij}_{\ \ \mu} &= \theta^i_{\ k} \omega^{kj}_{\ \ \mu} + \theta^j_{\ k} \omega^{ik}_{\ \ \mu} - \partial_\mu\theta^{ij} - \partial_\mu\xi^\rho \omega^{ij}_{\ \ \rho} - \xi^{\rho}\partial_{\rho}\omega^{ij}_{\ \ \mu}.
\end{aligned}
\end{align}
It may be easily checked that these transformations are indeed symmetries of the Einstein-Cartan action \eqref{EC3plus1} in 3+1 dimensions \cite{Blagojevic:2002du}.


We would now like to make a comparative remark on the structure of the Lagrangian generators in 2+1 and 3+1 dimensions. Let us consider the 2+1 dimensional generator $\bar{\zeta}^i_{\ \alpha k(0)}$ from \eqref{2plus1GenSet4}
\begin{align*}
\bar{\zeta}^i_{\ \alpha k(0)}(x,z) &= -\epsilon^i_{\ jk} \omega^j_{\ \alpha}(x) ~\delta({\bf x}-{\bf z}) - \delta^i_k~\partial_\alpha^{({\bf x})} \delta({\bf x}-{\bf z}).
\end{align*}
Multiplying appropriately with Levi-Civita symbols and using the map for $\omega^j_{\ \alpha}$ from \eqref{LiftMaps}, we get
\begin{align*}
\epsilon_i^{\ mn}\epsilon^k_{\ hp}\,\bar{\zeta}^i_{\ \alpha k(0)} = -\epsilon_i^{\ mn}\epsilon^k_{\ hp}\,\omega^i_{\ k\alpha}\,\delta(x-z) - \epsilon_i^{\ mn}\epsilon^i_{\ hp}~\delta_\alpha^{({\bf x})}\delta(x-z).
\end{align*}
Finally using the identity for contraction of Levi-Civita symbols \eqref{LeviCivitaProduct} and rearranging terms yield the following relation
\begin{align}
\label{ex d1}
\begin{aligned}
\bar{\zeta}^{mn}_{\ \ \ \alpha hp(0)} = &-\left[ \delta^m_h\omega^n_{\ p\alpha}-\delta^m_p\omega^n_{\  h\alpha}-\delta^n_h\omega^m_{\ p\alpha}+\delta^n_p\omega^m_{\ h\alpha} \right]\,\delta(x-z) \\
&- \left[ \delta^m_h\delta^n_p-\delta^m_p\delta^n_h \right]\,\partial_\alpha^{({\bf x})}\delta(x-z),
\end{aligned}
\end{align}
where we have defined the map
\begin{align}
\label{LagGenExMap}
\bar{\zeta}^{mn}_{\ \ \ \alpha hp(0)}(x,z) = -\frac{1}{2} ~\epsilon_i^{\ mn}\epsilon^k_{\ hp}\,\bar{\zeta}^i_{\ \alpha k(0)}(x,z).
\end{align}
Thus we have re-written the 2+1 dimensional generator $\bar{\zeta}^i_{\ \alpha k(0)}$ in terms of the original fields, getting rid of all duals. The object $\bar{\zeta}^{mn}_{\ \ \ \alpha hp(0)}$ defined in \eqref{ex d1}, however is functionally identical to the corresponding 3+1 dimensional generator \eqref{3plus1GenSet4}. Thus the map \eqref{LagGenExMap} expresses the 2+1 dimensional generator in a form that remains structurally the same even in 3+1 dimensions.

Similarly, all the other generators from 2+1 dimensions, can be stripped off the dual fields $\omega^j_{\ \sigma}$. These dual fields were defined for the special case of 2+1 dimensions. Once having removed them, and expressed all basic fields in terms of their dimension independent form, we see that the same generators also hold in 3+1 dimensions. Below, we list all the non-trivial maps, in the sense described above, between the Lagrangian generators.
\begin{align}
\label{GeneratorLiftMap}
\begin{aligned}
\bar{\rho}^{mn}_{\ \ \ 0\sigma(0)}(x,z) &= -\epsilon_i^{\ mn}\,\bar{\rho}^{i}_{\ 0\sigma(0)}(x,z)\\
\bar{\rho}^{mn}_{\ \ \alpha \sigma(0)}(x,z) &= -\epsilon_i^{\ mn}\,\bar{\rho}^{i}_{\alpha \sigma(0)}(x,z)\\
\bar{\rho}^{mn}_{\ \ \ 0\sigma(1)}(x,z) &= -\epsilon_i^{\ mn}\,\bar{\rho}^{i}_{\ 0\sigma(1)}(x,z)\\
\zeta^i_{\ \sigma mn(0)}(x,z) &= - \frac{1}{2}\epsilon^k_{\ mn}\,\zeta^i_{\ \sigma k(0)}(x,z)\\
\bar{\zeta}^{mn}_{\ \ \ \;\sigma hp(0)}(x,z) &= -\frac{1}{2} ~\epsilon_i^{\ mn}\epsilon^k_{\ hp}\,\bar{\zeta}^i_{\ \sigma k(0)}(x,z)\\
\bar{\zeta}^{mn}_{\ \ \ \;0hp(1)}(x,z) &= -\frac{1}{2} ~\epsilon_i^{\ mn}\epsilon^k_{\ hp}\,\bar{\zeta}^i_{\ 0k(1)}(x,z)
\end{aligned}
\end{align}
Observation of this structural similarity of the generators across dimensions, from 2+1 to 3+1, indicates that in higher dimensions, similar results are expected to hold.

\section{Conclusions}
\label{Sec:Conclu}

In this paper we have succeeded in systematically computing the generators of the PGT ({\it i.e.} Poincar\'{e} gauge theory) symmetries, which is a new result. The Lagrangian method of finding generators was employed, and in the case of first-order formulations of gravity, was seen to be much simpler than its Hamiltonian counterpart. To begin with, we took a 2+1 dimensional model of a PGT-invariant Lagrangian which has been of recent interest \cite{Banerjee:2009vf, Blagojevic:2009xw, Blagojevic:2003uc, Blagojevic:2004hj, Basu:2009dy} -- the 3D gravity model with torsion and a cosmological term. The starting gauge identities involving the Euler derivatives that were required for the Lagrangian analysis, were taken from a recent analysis \cite{Banerjee:2009vf}. The Lagrangian generators were subsequently computed and PGT symmetries were recovered using the same. We next repeated the procedure for 3+1 dimensions, where we took only the representative and most important Einstein-Cartan term in the action. We lifted the 2+1 dimensional gauge identities to 3+1 dimensions. The validity of the lifted gauge identities in 3+1 dimensions was explicitly checked. Then the same method was adopted to calculate the generators giving rise to PGT symmetries for this case. The Lagrangian generators themselves were also shown to preserve their structure across the 2+1 to 3+1 dimension transition. The PGT symmetries were shown to be consistent throughout this process as has been shown in Figure \ref{Fig1}.

Finding the generators that yield the PGT symmetries systematically, through a canonical method, was necessary to put this much studied symmetry at a firm level as it was not even known whether it was possible to generate the PGT symmetries by a canonical and completely off-shell process. A recent attempt \cite{Banerjee:2009vf} to generate the symmetries of a PGT invariant 2+1 dimensional Mielke-Baekler \cite{Mielke:1991nn} type model through pure Dirac Hamiltonian generator construction in a completely off-shell manner (following \cite{Banerjee:1999hu, Banerjee:1999yc}) had given symmetries which were distinctly different from PGT symmetries. The two sets only matched on-shell, {\it i.e.} after imposition of equations of motion. A similar conclusion was also noted in \cite{Blagojevic:2003uc, Blagojevic:2004hj} by following a modified Dirac Hamiltonian approach \cite{Castellani:1981us} where second class constraints were accounted by solving for the Lagrange multipliers rather than by using Dirac brackets. However, in this work, we have successfully shown that based on the Lagrangian generators we have constructed, one can now put this tetrad-connection formulation of symmetries on the same footing with the metric formulations \cite{Samanta:2007fk, Mukherjee:2007yi}, pertaining to the canonical method of constructing the symmetries.

\begin{appendix}

\renewcommand{\thesection}{Appendix \Alph{section}}		
\setcounter{section}{0}						

\section{Comments on general applications of Lagrangian generators}
\label{Sec:AppUsesLagGen}

\renewcommand{\theequation}{A.\arabic{equation}}		
\setcounter{equation}{0}  					

In this article, we demonstrated the role of Lagrangian generators in investigating gauge symmetries. In particular, the {\it off-shell} Poincar\'{e} gauge theory symmetries were reproduced through Lagrangian generators for the 2+1 and 3+1 dimensional Mielke Baekler type models of gravity. One might wonder about other possible applications of these generators. In this appendix we briefly comment on the issue.

As is well known, the canonical generators derived in the Hamiltonian formalism, apart from yielding the gauge symmetries, are also used to find the conserved charges and central terms in the Poisson algebra of spacetime symmetries \cite{Blagojevic:2006hh}. Since the Hamiltonian and Lagrangian formulations complement one another, it is expected that the Lagrangian generators will also have a similar, though not necessarily identical, role. We now elaborate on this and related points.

The crucial ingredient in abstracting the Lagrangian generators are the gauge identities. Construction of these identities can be made from physical considerations. However, there also exist systematic schemes for arriving at these gauge identities from algorithms employing Lagrangian constraints \cite{Chaichian:1994ug, Shirzad:1998af}. So, given a model with some Lagrangian, we can arrive at the gauge identities and the Lagrangian generators systematically. Now, some insight into these identities is gleaned from their connection with the Bianchi identities of a model \cite{Ortin:2004ms, Weinberg:1972}. In what follows, we adopt the Einstein-Hilbert action in 3+1 dimensions for the demonstration of this connection. The calculation follows \cite{Samanta:2007fk} closely.

The Einstein-Hilbert action in 3+1 dimensions is written in terms of the basic field $g_{\mu\nu}$ -- the metric -- as:
\begin{align}
\label{App action}
S=\int \textrm{d}^4x ~\sqrt{-g}\,g^{\mu\nu}R_{\mu\nu},
\end{align}
where the Ricci tensor $R_{\mu\nu}$ is defined in terms of the Christoffel connections $\Gamma^\rho_{\mu\nu}$ in the usual way as in Einstein general relativity:
\begin{align}
\label{App RicciChrist}
\begin{aligned}
R_{\mu\nu} &= \Gamma^\lambda_{\nu\mu,\,\lambda} - \Gamma^\lambda_{\lambda\mu,\,\nu} + \Gamma^\lambda_{\nu\mu}\,\Gamma^\sigma_{\sigma\lambda} - \Gamma^\sigma_{\lambda\mu}\,\Gamma^\lambda_{\nu\sigma}\\
\Gamma^\rho_{\ \mu\nu} &= \frac{1}{2} \, g^{\rho\lambda} \Big(\, g_{\lambda\nu,\,\mu} + g_{\mu\lambda,\,\nu} - g_{\mu\nu,\,\lambda} \,\Big).
\end{aligned}
\end{align}
Varying the action \eqref{App action} with respect to the metric $g_{\mu\nu}$ we get the Euler derivative $L^{\mu\nu}$
\begin{align}
\label{App VariedAction}
\delta S = \int \textrm{d}^4x ~L^{\mu\nu}\,\delta g_{\mu\nu}
\end{align}
where,
\begin{align}
\label{App EulerD}
L^{\mu\nu} = \sqrt{-g}\,G^{\mu\nu} = \sqrt{-g}\,\left( R^{\mu\nu} - \frac{1}{2}\,g^{\mu\nu}R \right).
\end{align}
Invariance of the action leads to the usual Einstein's equation $L^{\mu\nu}=0$. The gauge identity may be subsequently defined as
\begin{align}
\label{App GaugeId1}
\nabla_{\!\mu\,} L^\mu_{\ \nu} = 0
\end{align}
which may also be expressed as,
\begin{align}
\label{App GaugeId2}
\nabla_{\!\mu\,} G^\mu_{\ \nu} = 0.
\end{align}
Now, the Bianchi identity for Einstein general relativity is well known and is written in terms of the Riemann tensor $R_{\lambda\mu\nu\kappa}$ as:
\begin{align}
\label{App Bianchi}
\nabla_{\!\eta\,} R_{\lambda\mu\nu\kappa} + \nabla_{\!\nu\,} R_{\lambda\mu\kappa\eta} + \nabla_{\!\kappa\,} R_{\lambda\mu\eta\nu} = 0.
\end{align}
Contracting $\lambda$ with $\nu$ and $\mu$ with $\kappa$ in the above identity (the metricity condition $\nabla_{\!\rho\,}g_{\mu\nu} = 0$ holds in Einstein general relativity), we reproduce the gauge identity \eqref{App GaugeId2}. This immediately shows that the gauge identity is nothing but a suitably contracted form of the Bianchi identity in this model.

The gauge identity, or the contracted version of the Bianchi identity, plays a significant role in the obtention of the Noether central charges. As in the Hamiltonian description, here too surface terms are important. If these terms are not dropped, then \eqref{App VariedAction} takes the form,
\begin{align}
\label{App VariedActionWSur1}
\delta S = \int \textrm{d}^4x ~\sqrt{-g}\, \left[-G^{\mu\nu} \delta_{{\!}_\xi} g_{\mu\nu} + \nabla_{\!\rho}\left( 2\,g^{\mu\sigma,\rho\nu}\,\nabla_{\!\mu\,} \delta_{{\!}_\xi} g_{\sigma\nu} \right) \right].
\end{align}
Explicitly using $\delta_{{\!}_{\xi\,}} g_{\mu\nu} = \nabla_{\!(\mu\,}\xi_{\nu)}$ we obtain,
\begin{align}
\label{App VariedActionWSur2}
\delta S = \int \textrm{d}^4x ~\sqrt{-g}\,\Big[-2\,\left\lbrace \nabla_{\!\mu\,} G^{\mu\nu}\right\rbrace \xi_\nu + \nabla_{\!\rho} \left\lbrace 2\,G^{\rho\sigma}\xi_\sigma -4\, g^{\mu\sigma,\rho\nu}\,\nabla_{\!\mu} \nabla_{\!(\sigma\,}\xi_{\nu)} \right\rbrace \Big] = 0\,.
\end{align}
The first term in the integrand vanishes due to the gauge identity \eqref{App GaugeId2}. This also implies the vanishing of the second term in the integrand. Effectively, this leads to the covariant conservation of the Noether current,
\begin{align}
\label{App NoeCurr}
j_{\scriptscriptstyle N}^\rho(\xi) = 2\,R^{\,\rho\sigma}\,\xi_\sigma - 4\,g^{\mu\sigma,\rho\nu}\,\nabla_{\!\mu} \nabla_{\!(\sigma\,}\xi_{\nu)}.
\end{align}
Corresponding to each vector $\xi^\sigma$ it is now possible to construct a conserved Noether charge from \eqref{App NoeCurr}. This yields the standard Komar's integral in general relativity \cite{Note_Ortin}.

A comment on the surface terms might be useful. In the Hamiltonian approach, these terms are determined by requiring the functional differentiability of the generators. The corresponding criterion in the present Lagrangian formulation is to retain all surface terms in the variation of the action under a general coordinate transformation, eventually leading to the gauge identity. This is clearly manifested in \eqref{App VariedActionWSur2} where the first term in the integrand yields the gauge identity while the second is the cherished surface term.

We thus observe how the gauge identity, which is directly connected with the Lagrangian generators, leads to conserved Noether charges. Also, the complementary aspects of Lagrangian and Hamiltonian generators, vis-$\grave{\text{a}}$-vis the construction of conserved charges gets illuminated.

\end{appendix}

\end{document}